\documentstyle[11pt,aaspp]{article} 

\slugcomment{COBE Preprint 96-01, Submitted To ApJ Letters}

\lefthead{Bennett et al.}
\righthead{Four Year COBE DMR Results}

\def\check_mode#1{\ifmmode{#1}\else{$#1$}\fi}
\def\deg    {\check_mode{^\circ\ }}

\def\lsim {\check_mode{_<\atop^{\sim}}}

\def\lt     {\check_mode{<}}

\begin{document}

\title{4-Year COBE\altaffilmark{1}
    DMR Cosmic Microwave Background Observations:\\Maps and Basic Results\\}

\author{C. L. Bennett\altaffilmark{2,3},
        A. Banday\altaffilmark{4}, 
        K. M. G\'{o}rski\altaffilmark{4}, 
        G. Hinshaw\altaffilmark{4}, 
        P. Jackson\altaffilmark{4}, 
        P. Keegstra\altaffilmark{4}, 
        A. Kogut\altaffilmark{4},
        G. F. Smoot\altaffilmark{5} 
        D. T. Wilkinson\altaffilmark{6} 
     \& E. L. Wright\altaffilmark{7}}

\altaffiltext{1}{NASA/GSFC is responsible for the design, 
    development, and operations of the {\it COBE}.  Scientific guidance is 
    provided by the {\it COBE} Science Working Group.  GSFC is also
    responsible for the development of the analysis software
    and the delivery of the mission data sets.}
\altaffiltext{2}{Code 685, Laboratory for Astronomy \& Solar
       Physics, Goddard Space Flight Center, Greenbelt, MD 20771.}
\altaffiltext{3}{e-mail address: bennett@stars.gsfc.nasa.gov}
\altaffiltext{4}{Hughes STX Corporation, Code 685, 
Laboratory for Astronomy \& Solar
       Physics, Goddard Space Flight Center, Greenbelt, MD 20771.}
\altaffiltext{5}{LBL and UC Berkeley, Berkeley, CA 94720.}
\altaffiltext{6}{Physics Dept., Jadwin Hall, Princeton University, Princeton,
NJ 08544-1001.} 
\altaffiltext{7}{UCLA Astronomy, P.O. Box 951562, Los Angeles, CA 90095-1562.}

\begin{abstract}
The cosmic microwave background radiation provides unique 
constraints on cosmological models.  In this {\it Letter} we present a summary
of the spatial properties of the cosmic microwave 
background radiation based on the full 4 years of {\it COBE} DMR observations,
as detailed in a set of companion {\it Letters}.
The anisotropy is consistent with a scale-invariant power law model and
Gaussian statistics.  With full use of the multi-frequency 4-year 
DMR data, including our estimate of the effects of Galactic emission, we find 
a power-law spectral index of $n=1.2\pm 0.3$ and a quadrupole normalization 
$Q_{rms-PS}=15.3^{+3.8}_{-2.8}$ $\mu$K. For $n=1$ the best-fit normalization is
$Q_{rms-PS}\vert_{n=1}=18\pm 1.6$ $\mu$K. These values are consistent with
both our previous 1-year and 2-year results.  The results include use of the 
$\ell=2$ quadrupole term; exclusion of this term gives consistent results, but
with larger uncertainties.  The 4-year sky maps,
presented in this {\it Letter}, 
portray an accurate overall visual impression of the 
anisotropy since the signal-to-noise ratio is $\sim2$ per $10^\circ$ sky map
patch.  The improved signal-to-noise ratio of the 4-year maps also allows 
for improvements in Galactic modeling and limits on non-Gaussian statistics.
\end{abstract}


\section{INTRODUCTION}

NASA's {\it COBE} Differential Microwave Radiometers (DMR) experiment
(\cite{Smoot90,Bennett91}) discovered cosmic microwave background (CMB)
anisotropies based on its first year of data 
(\cite{Smoot92,Bennett92,Wright92}).  
The CMB temperature fluctuations were measured at an angular 
resolution of $7^\circ$ at frequencies of 31.5, 53, and 90 GHz.  These 
results were supported by a detailed examination of the DMR calibration and 
its uncertainties (\cite{Bennett91}) and a detailed treatment of the upper 
limits on residual systematic errors (\cite{Kogut92}).  Bennett et al. (1992) 
showed that spatially correlated Galactic free-free and dust emission could 
not mimic the frequency spectrum nor the spatial distribution of the observed 
fluctuations.  
Bennett et al. (1993) also showed that the pattern of fluctuations 
does not spatially correlate with known extragalactic source distributions.  
Confirmation of the {\it COBE} results was attained by the positive 
cross-correlation between the {\it COBE} data and data from balloon-borne 
observations at a shorter wavelength (\cite{Ganga93}).  Bennett et al. (1994) 
reported the results from analyses of two years of DMR data.  The 
results from the two year data were consistent with those from the first year 
alone.  In this {\it Letter} we summarize the results and cosmological 
implications obtained from the full {\it COBE} DMR 4-year data, and provide 
references to further detailed reports of our analyses.

Primordial gravitational potential fluctuations were predicted to have an equal
$rms$ amplitude on all scales (\cite{Peebles70,Harrison70,Zeldovich72}).  
This corresponds to a matter fluctuation power-law spectrum, $P(k)\propto k^n$,
where $k$ is the comoving wavenumber, with $n=1$.  Such a spectrum is also
a natural consequence of inflationary models.  If the effects of a standard cold
dark matter model are included, 
{\it COBE} DMR should find $n_{eff}\approx 1.1$ for 
a Peebles-Harrison-Zeldovich $n=1$ universe.  The power spectrum of the {\it
COBE} DMR data is consistent with this.

A cosmological model does not predict the exact CMB temperature that would be 
observed in our sky, but rather predicts a statistical distribution of
anisotropy parameters, such as spherical harmonic amplitudes.  In the context 
of such models, the true CMB temperature observed 
in our sky is only a single realization from a statistical distribution.  
Thus, in addition to experimental uncertainties, we account for 
{\it cosmic variance} uncertainties in our analyses of the DMR maps.  
For a spherical harmonic temperature expansion 
$T(\theta,\phi)=\sum_{\ell m} a_{\ell m}Y_{\ell m}(\theta,\phi)$,
cosmic variance is approximately expressed
as $\sigma(C_\ell)/C_\ell \approx \sqrt{2/(2\ell+1)}$
where $C_\ell=<\vert a_{\ell m}\vert^2>$.
Cosmic variance exists independent of the quality of the experiment.  The
power spectrum from the 4-year DMR map is cosmic variance limited for 
$\ell \lsim 20$.

This {\it Letter} includes a summary of the key results of a set of
detailed DMR 4-year analysis papers (Banday et al. 1996a, 1996b; 
Gorski et al. 1996; 
Kogut et al. 1996a, 1996b, 1996c; Hinshaw et al. 1996a, 1996b; 
Wright et al. 1996).

\section{OBSERVATIONS}

DMR consists of 6 differential microwave radiometers: 2 nearly independent 
channels, labeled A and B, at frequencies 31.5, 53, and 90 GHz (wavelength 
9.5, 5.7, and 3.3 mm).  Each radiometer measures the difference in power
between two 7\deg fields of view separated by 60\deg, 30\deg to either side 
of the spacecraft spin axis (\cite{Smoot90}).  {\it COBE} was launched from 
Vandenberg Air Force Base on 18 November 1989 into a 900 km, 99\deg 
inclination circular orbit, which precesses to follow the terminator.  Attitude 
control keeps the spacecraft pointed away from the Earth and nearly 
perpendicular to the Sun so that solar radiation never directly illuminates 
the aperture plane.  The combined motions of the spacecraft spin (75 s period),
orbit (103 m period), and orbital precession ($\sim 1\deg$ per day) allow 
each sky position to be compared to all others through a highly redundant set 
of temperature difference measurements spaced 60\deg apart.
The on-board processor integrates the differential signal from each channel 
for 0.5 s, and records the digitized differences for daily playback to a 
ground station.  

Ground data analysis consists of calibration, extensive systematic error
analyses, and conversion of time-ordered-data to sky maps (Kogut et al. 1996a).
The DMR time-ordered-data include systematic effects such as emission from the
Earth and Moon, the instrument's response to thermal changes, and the
instrument's response to the Earth's magnetic field. The largest detected
effects do not contribute significantly to the DMR maps: they are either on
time scales long compared to the spacecraft spin sampling (e.g. thermal gain
drifts) or have time dependence inconsistent with emission fixed on the
celestial sphere (e.g. magnetic effects). Detected and potential systematic
effects were quantitatively analyzed in detail by Kogut et al. (1996a). 
Data with the worst systematic contamination 
(lunar emission, terrestrial emission, and thermal gain changes)
were not used in the map making process 
and constitute less than 10\% of the data in the 53 and 90 GHz channels.
The remaining data were corrected using models of each effect.
The data editing and correction parameters were conservatively chosen 
so that systematic artifacts, after correction, are less than 6 $\mu$K (95\%
confidence upper limit) in the final DMR map in the worst channel.  This is
significantly less than the levels of the noise and celestial signals.

We subtract a dipole (with Cartesian 
components [X,Y,Z] = [-0.2173, -2.2451, +2.4853] mK
thermodynamic temperature in Galactic coordinates) from the time-ordered 
differential data prior to forming the 4-year sky maps to reduce spatial
gradients within a single pixel.  A small residual dipole 
remains in the maps from a combination of CMB and Galactic emission.
Fig. 1 shows the full sky maps at each frequency, after averaging the A and B 
channels, removing the CMB dipole, and smoothing to 10\deg effective resolution.
Table \ref{noise_table} shows the instrument noise per half-second observation.
The mean signal-to-noise ratios in the 10\deg smoothed 
maps are $\sim$ 0.5, 1.5, 
and 1.0 for 31, 53, and 90 GHz, respectively.  
For a multi-frequency co-added map
the signal-to-noise ratio is $\sim$ 2.  This signal-to-noise level is adequate
to portray an accurate overall visual impression of the anisotropy.  This is
illustrated in Fig. 2, where simulated data are shown in combination with the
noise appropriate to 1-, 2-, and 4-years of DMR 53 GHz observations.

Given the sensitivity of the 4-year DMR maps we have chosen to extend
the cut made in our previous analyses to exclude additional Galactic
emission.  Along with the previous $\vert b\vert<20\deg$ exclusion
zone, we use the {\it COBE} DIRBE 140 $\mu$m map as a guide to cut
additional Galactic emission features. The full sky DMR maps contain
6144 pixels.  An optimum Galactic cut maximizes the number of remaining
pixels while minimizing the Galactic contamination.  Fig. 3 shows the
residual Galactic signal as a function of the number of usable pixels
after a cut is applied based on the 140 $\mu$m DIRBE intensity.  Our
cut leaves 3881 pixels (in Galactic pixelization) while eliminating
the strongest $\vert b\vert>20\deg$ Galactic emission. Moderate
changes to this custom cut will cause derived CMB parameters to change
somewhat, but this is consistent with the data sampling differences of
real CMB anisotropy features and not necessarily Galactic
contamination.  Likewise, derived CMB parameters also vary by the
expected amount when the maps are made in ecliptic rather than
Galactic coordinates since about 1/2 of the noise is re-binned.

Kogut et al. (1996b) examine the Galactic contamination of the surviving high
Galactic latitude regions of the DMR maps after the `custom cut' (described
above) is applied. No significant cross-correlation is found between the DMR
maps and either the 408 MHz synchrotron map or the synchrotron map derived from
a magnetic field model (Bennett et al. 1992). This places an
upper limit $T_{\rm synch} < 11 ~\mu$K (95\% confidence) on synchrotron 
emission at 31 GHz. 

A significant correlation is found between the DMR maps and the dust-dominated 
DIRBE 140 $\mu$m map, with frequency dependence consistent with a superposition
of dust and free-free emission.  This corresponds to 
a 7\deg rms free-free emission component of $7.1\pm 1.7$ $\mu$K at 53 GHz
and a dust component of $2.7\pm 1.3$ $\mu$K at 53 GHz.  Since this emission
is uncorrelated with CMB anisotropies it constitutes $<10$\% of the CMB power.
The amplitude of the correlated free-free component at 53 GHz
agrees with a noiser estimate of free-free emission 
derived from a linear combination of DMR data which includes 
{\it all} emission with free-free spectral dependence.
The combined dust and free-free emission contribute $10 \pm 4 ~\mu$K {\it rms}
at both 53 and 90 GHz, well below the 30 $\mu$K cosmic signal.
These Galactic signal analyses are consistent with the fact that the fitted 
cosmological parameters are nearly unaffected by removal of 
modeled Galactic signals (G\'{o}rski et al.\ 1996, Hinshaw et al.\ 1996),
with the notable exception of the quadrupole,
which has significant Galactic contamination (Kogut et al.\ 1996b).  A search
by Banday et al. (1996a) finds no evidence for significant extragalactic 
contamination of the DMR maps.

\section{INTERPRETATION}

{\bf Monopole $\bf \ell=0$:}  Despite the fact that the DMR is a differential 
instrument, the known motion 
of the {\it COBE} spacecraft about the Earth and the motion of the
Earth about the Solar System barycenter allows a determination of the
CMB monopole temperature from the DMR data.  The CMB at millimeter
wavelengths is well described by a blackbody spectrum
(\cite{Mather94,Fixsen96}).  The Doppler effect from the combined
spacecraft and Earth orbital motions creates a dipole signal
$T(\theta) = T_0 [ 1 + \beta \cos(\theta) + O(\beta^{2}) ]$, where
$\beta = v/c$ and $\theta$ is the angle relative to the time-dependent
velocity vector.  The satellite and Earth orbital motions are well
known and change in a regular fashion, allowing their Doppler signal
to be separated from fixed celestial signals.  We fit the time-ordered
data to the Doppler dipole and recover a value for the CMB monopole
temperature, $T_0 = 2.725 \pm 0.020$ K (\cite{Kogut96a}).

{\bf Dipole $\bf \ell=1$:}
The CMB anisotropy is dominated by a dipole term usually attributed to the 
motion of the Solar System with respect to the CMB rest frame, as seen in Fig.
4.  A precise determination of the dipole must account for Galactic emission 
and the aliasing of power from higher multipole orders once pixels near the 
Galactic plane are 
discarded.  We account for Galactic emission using a linear combination of 
the DMR maps or by cross-correlating the DMR maps with template sky maps 
dominated by Galactic emission (\cite{Kogut96b}).  We fit the high-latitude 
portion of the sky for a dipole with a CMB frequency spectrum using a 
pixel-based likelihood analysis (\cite{Hinshaw96a}).  Accounting for the 
smoothing by the DMR beam and map pixelization, the CMB dipole has amplitude 
$3.353 \pm 0.024$ mK toward Galactic coordinates 
$(l,b) = (264\fdg26 \pm 0\fdg33, 48\fdg22 \pm 0\fdg13)$,
or equatorial coordinates
$(\alpha, \delta) = (11^{\rm h} 12\fm2 \pm 0\fm8, -7\fdg06 \pm 0\fdg16)$
epoch J2000.

{\bf Quadrupole $\bf \ell=2$:}
On the largest angular scales (e.g., quadrupole), Galactic emission is
comparable in amplitude to the anisotropy in the CMB.  We use a
likelihood analysis to fit the high-latitude portion of the DMR maps
for Galactic emission traced by synchrotron- and dust-dominated
surveys and a quadrupole anisotropy with a thermodynamic frequency
spectrum (\cite{Kogut96b,Hinshaw96a}). After correcting for the
positive bias from instrument noise and aliasing, the CMB quadrupole
amplitude observed at high latitude is $Q_{rms} = 10.7\pm 3.6\pm 7.1$
$\mu$K, where the quoted errors reflect the 68\% confidence
uncertainties from random statistical errors and Galactic modeling
errors, respectively.  The observed quadrupole amplitude, $Q_{rms}$,
has a lower value than the quadrupole expected from a fit to the
entire power spectrum, $Q_{rms-PS}$, but whether this is a chance
result of cosmic variance or reflects the cosmology of the universe
can not be determined from {\it COBE} data.  The 68\% confidence
interval for the quadrupole amplitude, 6 $\mu$K $\le$ $Q_{rms}$ $\le$
17 $\mu$K, is consistent with the quadrupole normalization of the full
power spectrum power-law fit (discussed below): $Q_{rms-PS} =
15.3^{+3.8}_{-2.8}$ $\mu$K.

{\bf Power spectrum $\bf \ell \ge 2$:}
The simplest probe of the angular power spectrum of the anisotropy is its 
Legendre transform, the 2-point correlation function.  The 2-point correlation
function of 
the 4-year maps is analyzed by Hinshaw et al. (1996b), where it is shown that 
the 2-point data are consistent from channel to channel and frequency 
to frequency.  The data are robust with respect to the angular power 
spectrum.  As in Bennett et al. (1994), we use a Monte Carlo-based Gaussian 
likelihood analysis to infer the most-likely quadrupole normalization for a 
scale-invariant ($n=1$) power-law spectrum.  The results are summarized in 
Table \ref{Qn_fit} where we also include the results of 3 additional, 
independent power spectrum analyses, discussed below.  The normalization 
inferred from the 2-point function is now in better agreement with other 
determinations than was the case with the 2-year data.  The change is due to 
data selection: with the 2-year data, we only analyzed the 53 $\times$
90 GHz cross-correlation function; with the 4-year data we have
analyzed many more data combinations, including the auto-correlation
of a co-added, multi-frequency map.  This latter combination is more
comparable to the data analyzed by other methods, and the 2-point
analysis yields consistent results in that case.  The combined 31, 53
and 90 GHz CMB rms is 29$\pm$1 $\mu$K in the 10\deg smoothed map
(Banday et al. 1996b), consistent with the level determined by the
2-point results.

It is also possible to analyze the power spectrum directly in terms of
spherical harmonics.  However, there is considerable subtlety in this
because the removal of the Galactic plane renders the harmonics
non-orthonormal, producing strong correlations among the fitted
amplitudes. Wright et al. (1996) has solved for an angular power
spectrum by modifying and applying the technique described by Peebles
(1973) and Hauser \& Peebles (1973) for data on the cut sphere.  They
compute a Gaussian likelihood on these data and calibrate their
results with Monte Carlo simulations.  G\'{o}rski et al. (1996)
explicitly construct orthonormal functions on the cut sphere and
decompose the anisotropy data with respect to these modes.  They form
and evaluate an exact Gaussian likelihood directly in terms of this
mode decomposition.  The results of these analyses are summarized in
Table \ref{Qn_fit}.  Further details, including results from other
data combinations are given in the respective papers.

Hinshaw et al. (1996a) evaluate a Gaussian likelihood directly in
terms of a full pixel-pixel covariance matrix, a technique applied to
the 2-year data by Tegmark \& Bunn (1995).  The results of the
power-law spectrum fits are summarized in Table \ref{Qn_fit}.  Hinshaw
et al. (1996a) also analyze the quadrupole anisotropy separately from
the higher-order modes, to complement the analysis of Kogut et al.
(1996b).  They compute a likelihood for the quadrupole mode $C_2$,
nearly independent of higher-order power, and show that it peaks
between 6 and 10 $\mu$K, depending on Galactic model, but that its
distribution is so wide that it is easily consistent with
$15.3^{+3.8}_{-2.8}$ $\mu$K, the value derived using the full power
spectrum.

{\bf Tests for Gaussian Statistics:}
It is important to determine whether the primordial fluctuations are
Gaussian. The probability distribution of temperature residuals should
be close to Gaussian if the sky variance is Gaussian and the receiver
noise is Gaussian.  The receiver noise varies somewhat from pixel to
pixel because the observation times are not all the same, but when
this is taken into account the data appear Gaussian (\cite{Smoot94}). 
There is no evidence for an excess of large deviations, as would be
expected if there were an unknown population of point sources. A
search for point sources in the 2-year maps was negative
(\cite{Kogut94}). Given the large beam of the instrument and the
variance of both cosmic signals and receiver noise, it is still
possible for interesting signals to be hidden in the data.

Kogut et al. (1996c) compare the 4-year DMR maps to Monte Carlo simulations of 
Gaussian power-law CMB anisotropy.  The 3-point correlation function, 
the 2-point correlation of temperature extrema, and the topological genus are 
all in excellent agreement with the hypothesis that the CMB anisotropy on 
angular scales of 7\deg or larger represents a random-phase Gaussian field.
A likelihood comparison of the DMR maps against non-Gaussian $\chi^2_N$ toy 
models tests the alternate hypothesis that the CMB is a random realization
of a field whose spherical harmonic coefficients $a_{\ell m}$ are drawn from 
a $\chi^2$ distribution with $N$ degrees of freedom.  Not only do Gaussian 
power-law models provide an adequate description of the large-scale CMB 
anisotropy, but non-Gaussian models with $1 \lt N \lt 60$ are five times less 
likely to describe the true statistical distribution than the exact Gaussian 
model.

\section{SUMMARY OF 4-YEAR {\it COBE} DMR CMB MEASUREMENTS}

(1) The full 4-year set of {\it COBE} DMR observations is analyzed and
full sky maps are presented.  The typical signal-to-noise ratio in a
10\deg smoothed map is $\sim 2$ in the frequency-averaged map, enough
to provide a visual impression of the anisotropy.

(2) We derive a CMB monopole temperature from DMR (despite its being a
differential instrument) of $T_0 = 2.725 \pm 0.020$ K (Kogut et al. 1996a).  
This is in excellent agreement
with the {\it COBE} FIRAS precision measurement of the spectrum of the CMB,
$T_0 = 2.728 \pm 0.002$ K (Fixsen et al. 1996). 

(3) The CMB dipole from DMR has amplitude $3.353 \pm 0.024$ mK toward Galactic 
coordinates 
$(l,b) = (264\fdg26 \pm 0\fdg33, 48\fdg22 \pm 0\fdg13)$, 
or equatorial coordinates 
$(\alpha, \delta) = (11^{\rm h} 12\fm2 \pm 0\fm8, -7\fdg06 \pm 0\fdg16)$ 
epoch J2000.  This is consistent with the dipole
amplitude and direction derived by {\it COBE} FIRAS (Fixsen et al.
1996).

(4) The 95\% confidence interval for the observed $\ell=2$ quadrupole
amplitude is 4 $\mu$K $\le$ $Q_{rms}$ $\le$ 28 $\mu$K.  This is
consistent with the value predicted by a power-law fit to the power
spectrum: $Q_{rms-PS} = 15.3^{+3.8}_{-2.8}$ $\mu$K (Kogut et al.
1996b; Hinshaw et al. 1996a).

(5) The power spectrum of large angular scale CMB measurements are
consistent with an $n=1$ power-law (G\'{o}rski et al. 1996, Hinshaw et
al. 1996a, Wright et al. 1996).  If the effects of a standard cold
dark matter model are included, {\it COBE} DMR should find
$n_{eff}\approx 1.1$ for a $n=1$ universe.) With full use of the
multi-frequency 4-year DMR data, including our estimate of the effects
of Galactic emission, we find a power-law spectral index of $n=1.2\pm
0.3$ and a quadrupole normalization $Q_{rms-PS}=15.3^{+3.8}_{-2.8}$
$\mu$K. For $n=1$ the best-fit normalization is
$Q_{rms-PS}\vert_{n=1}=18\pm 1.6$ $\mu$K.  Differences in the derived
values of $Q$ and $n$ between various analyses of DMR data are much
more dependent on the detailed data selection effects than on the
analysis technique.

(6) The DMR anisotropy data are consistent with Gaussian statistics.
Statistical tests prefer Gaussian over other toy statistical models
by a factor of $\sim 5$ (Kogut et al. 1996c).

The DMR time-ordered data, map data, and ancillary data sets (including our
custom Galactic cut) are publicly available through the National Space Science 
Data Center (NSSDC) at http://www.gsfc.nasa.gov/astro/cobe/cobe\_home.html. 
We gratefully acknowledge the {\it COBE} support provided
by the NASA Office of Space Sciences (OSS).  We also acknowledge the
contributions of J. Aymon, V. Kumar, R. Kummerer, C. Lineweaver, J. Santana,
and L. Tenorio.

\begin{planotable}{cccc}
\tablewidth{3.in}
\tablecaption{DMR Instrument Noise in Antenna Temperature}
\tablehead{ \colhead{}                      &
            \colhead{}                      &
            \colhead{Noise per 0.5 sec}     \\
            \colhead{}                      &
            \colhead{Channel}               &
            \colhead{Observation}           &
            \colhead{}                      \\
            \colhead{}                      &
            \colhead{}                      &
            \colhead{(mK)}                  &
            \colhead{}                      }
\startdata
&31A & 58.27 &\nl
&31B & 58.35 &\nl
&53A & 23.13 &\nl
&53B & 27.12 &\nl
&90A & 39.10 &\nl
&90B & 30.76 &\nl
\label{noise_table}
\end{planotable}

\begin{planotable}{lccc}
\tablewidth{4.in}
\tablecaption{Summary of DMR 4-Year Power Spectrum Results}
\tablehead{ \colhead{Statistic}                          &
            \colhead{$n$\tablenotemark{a}}               &
            \colhead{$Q_{rms-PS}$\tablenotemark{b}}      &
            \colhead{$Q_{rms-PS|n=1}$\tablenotemark{c}}  \nl
            \colhead{}                                   &
            \colhead{}                                   &
            \colhead{($\mu$K)}                           &
            \colhead{($\mu$K)}                           }
\startdata
\multicolumn{4}{c}{No Galaxy Correction\tablenotemark{d}}\nl
2-point function\tablenotemark{e}   & ---     & ---     & $18.6^{+1.4}_{-1.4}$ \nl
Orthog. functions\tablenotemark{f}    & $1.23^{+0.23}_{-0.29}$ &
                     $15.3^{+3.9}_{-2.6}$ & $18.3^{+1.3}_{-1.2}$ \nl
Pixel temps\tablenotemark{g}       & $1.25^{+0.26}_{-0.29}$ &
                     $15.4^{+3.9}_{-2.9}$ & $18.4^{+1.4}_{-1.3}$ \nl
Hauser-Peebles\tablenotemark{h}& $1.30^{+0.30}_{-0.34}$ &
                     ---                & $17.7^{+1.4}_{-1.8}$ \nl
\multicolumn{4}{c}{DIRBE-Template Galaxy Correction\tablenotemark{i}}\nl
2-point function\tablenotemark{e}   & ---     & ---     & $17.5^{+1.4}_{-1.4}$ \nl
Orthog. functions\tablenotemark{f}    & $1.21^{+0.24}_{-0.28}$ &
                     $15.2^{+3.7}_{-2.6}$ & $17.7^{+1.3}_{-1.2}$ \nl
Pixel temps\tablenotemark{g}       & $1.23^{+0.26}_{-0.27}$ &
                     $15.2^{+3.6}_{-2.8}$ & $17.8^{+1.3}_{-1.3}$ \nl
Hauser-Peebles\tablenotemark{h}& ---  & --- & ---  \nl
\multicolumn{4}{c}{Internal Combination Galaxy Correction\tablenotemark{j}}\nl
2-point function\tablenotemark{e}   & ---     & ---     & $16.7^{+2.0}_{-2.0}$ \nl
Orthog. functions\tablenotemark{f}    & $1.11^{+0.38}_{-0.42}$ &
                     $16.3^{+5.2}_{-3.7}$ & $17.4^{+1.8}_{-1.7}$ \nl
Pixel temps\tablenotemark{g}       & $1.00^{+0.40}_{-0.43}$ &
                     $17.2^{+5.6}_{-4.0}$ & $17.2^{+1.9}_{-1.7}$ \nl
Hauser-Peebles\tablenotemark{h}& $1.62^{+0.44}_{-0.50}$ &
                     ---                 & $19.6^{+2.5}_{-2.5}$\nl
\tablenotetext{a}{Mode and 68\% confidence range of the projection of 
the 2-dimensional likelihood, $L(Q,n)$, on $n$}
\tablenotetext{b}{Mode and 68\% confidence range of the projection of 
the 2-dimensional likelihood, $L(Q,n)$, on $Q$}
\tablenotetext{c}{Mode and 68\% confidence range of the slice of the 
2-dimensional likelihood, $L(Q,n)$, at $n=1$}
\tablenotetext{d}{Formed from the weighted average of all 6 channels}
\tablenotetext{e}{Hinshaw et al. 1996b}  
\tablenotetext{f}{G\'orski et al. 1996}
\tablenotetext{g}{Hinshaw et al. 1996a}  
\tablenotetext{h}{Wright et al. 1996; in this case data selection differs
slightly, as described in the reference}  
\tablenotetext{i}{Formed from the weighted average of all 6 channels
with the best-fit Galactic template maps subtracted (Kogut et al. 1996b)}
\tablenotetext{j}{Formed from a linear combination of all 6 channel maps
that cancels free-free emission (Kogut et al. 1996)}
\label{Qn_fit}
\end{planotable}

FIGURE CAPTIONS

Figure 1: [color plate] Full sky Mollweide projections of the 31, 53, and 90 GHz
4-year maps formed by averaging the A and B channels at each frequency.  After 
the mean offset and the dipole are removed, the bright
(red) emission from the Galaxy dominates the central band of the maps.

Figure 2: [color plate] ({\it top left}) 
Full sky simulation of an anisotropy map 
smoothed to 10\deg resolution with an $n=1$ Peebles-Harrison-Zeldovich power 
spectrum with $Q_{rms-PS}=18$
$\mu$K.  DMR instrument noise is added to the sky simulation corresponding to: 
({\it top right}) 1-year; ({\it bottom right}) 2-years; ({\it bottom left}) 
4-years of observations.  
Note the good visual agreement between the `true' sky on the top
left and the 4-year simulated DMR result on the bottom left.

Figure 3: The estimated residual Galactic signal at 53 GHz as a function of the
number of surviving map pixels after the application of an intensity cut that
is based on the DIRBE 140 $\mu$m map.  The conversion from DIRBE intensity to 
DMR 53 GHz intensity is from a correlation study (Kogut et al. 1996b).  
A minimum $\vert b\vert <$ 20\deg cut is always applied, 
leaving at most 4016 pixels. The
arrows indicate the intensity cut that we chose to maximize the remaining
number of pixels while minimizing the Galactic contamination. This leaves 3881
map pixels for CMB analysis. A simple $\vert b\vert <$ 30\deg cut produces the
peak-to-peak intensity and rms shown by the '+' symbols, and is obviously
less efficient than the custom cut.

Figure 4: [color plate] 
({\it top}) Full sky Mollweide projection of the 4-year 53 GHz DMR map,
including the dipole.
({\it middle}) Full sky Mollweide projection of the 4-year 53 GHz DMR map, 
excluding the dipole.
({\it bottom}) Full sky Mollweide projection of the 4-year DMR map, 
excluding the dipole, 
using data from 31, 53, and 90 GHz with modeled Galactic emission
removed, and the Galactic custom cut applied.

\end{document}